# Discovery of Hidden Classes of Layered Electrides by Extensive High-throughput Materials Screening

Jun Zhou,[†] Lei Shen,[§] Ming Yang,[‡,∥] Haixia Cheng,[†,⊥] Weilong Kong,[†] and Yuan Ping Feng*[,†,‡]

[†]Department of Physics, National University of Singapore, Singapore 117411, Singapore.

[‡]Centre for Advanced 2D Materials and Graphene Research Centre, National University of Singapore, Singapore 117546, Singapore.

[§]Mechanical Engineering, National University of Singapore, Singapore 117575, Singapore.

[∥]Institute of Materials Research and Engineering, A*-STAR, 2 Fusionopolis Way, Singapore 138634, Singapore.

[⊥]Department of Physics, University of Science and Technology Beijing, Beijing 100083, China

**ABSTRACT:** Despite their extraordinary properties, electrides are still a relatively unexplored class of materials with only a few compounds grown experimentally. Especially for layered electrides, the current researches mainly focus on several isostructures of $Ca_2N$ with similar interlayer two-dimensional (2D) anionic electrons. An extensive screening for different layered electrides is still missing. Here, by screening materials with anionic electrons for the structures in Materials Project, we uncover 12 existing materials as new layered electrides. Remarkably, these layered electrides demonstrate completely different properties from $Ca_2N$. For example, unusual fully spin-polarized zero-dimensional (0D) anionic electrons are shown in metal halides with $MoS_2$-like structures; unique one-dimensional (1D) anionic electrons are confined within the tubes of the quasi-1D structures; a coexistence of magnetic and non-magnetic anionic electrons is found in ZrCl-like structures and a new ternary $Ba_2LiN$ with both 0D and 1D anionic electrons. These materials not only significantly increase the pool of experimentally synthesizable layered electrides but also are promising to be exfoliated into advanced 2D materials.

## INTRODUCTION

Electrides are a unique class of compounds in which some electrons are neither fully delocalized like in metals nor bind to any particular atom, molecule or bond, but sit in lattice interstitial spaces, functioning as anions.[1,2] Such anionic electrons can be highly active, making electrides chemically and thermally unstable. Despite many endeavors since the first crystalline electride grown three decades ago,[3] only a few electrides have been synthesized.[4-13] The first room-temperature stable inorganic electride was reported in 2003 by removing oxygen ions from the crystallographic cages of $12CaO·7Al_2O_3$, yielding $[Ca_{24}Al_{28}O_{64}]^{4+}(4e^-)$.[5] The high density of anionic electrons with loosely bounded nature makes $[Ca_{24}Al_{28}O_{64}]^{4+}(4e^-)$ promising for a plethora of applications, such as efficient thermionic emitters,[14] electron-injection electrode for organic light-emitting diodes,[15,16] and high performance catalysts.[17-20]

Recently, layered electrides are attracting increasing research interest both for fundamental physics and applications. The seminal example is the layered dicalcium nitride ($Ca_2N$) with an Anti-$CdCl_2$ structure (R-3m), which can be taken as $[Ca_2N]^+(e^-)$. Remarkably, the excess electrons from counting oxidation numbers were confined in the space between the $[Ca_2N]^+$ layers, forming dense two-dimensional (2D) electron layer,[6,21] promising for electron dopant,[22-26] batteries,[27,28] and plasmonic device applications.[29-31] Meanwhile, monolayer $Ca_2N$ preserving its unique two-dimensional electron layers in interstitial space was predicted to be stable theoretically and subsequently was grown experimentally.[32,33] This proves the possibility of exfoliation of layered electrides into advanced 2D materials, appealing for applications such as nanoelectronics.[34,35]

Motivated by $Ca_2N$, several theoretical calculations have been performed to design new layered inorganic electrides. A global structure optimization method with further phonon spectrum and molecular dynamic simulations were applied to find new thermodynamically stable electrides.[36] Six binary materials with hexagonal crystal structure for $A_2B$ (A=alkaline-earth metals, B=VA or VIA non-metals) and tetragonal crystal structure for AB (A=alkaline-earth metals, B=halogen) have been theoretically shown to be thermodynamically stable layered electrides. Another study based on the interstitial electron localization instead of total energy as the global variable

function has generated 89 new inorganic binary electrides.[37]

Considering the highly unstable nature of electrides, theoretical designs may suffer from the question mark if they can be grown experimentally. For example, the calculated thermodynamically stable Sr$_2$P predicted in the previous study was later found to be metastable and a different structure was identified as the most stable phase by a genetic algorism search with first-principles calculations, which was further validated by experiments.[38] This demonstrates the particularly important role of experimental validation and support in discussion on hidden electrides. Database screening to identify candidates from grown compounds is a shortcut to find stable electrides. In fact, the recently reported layered electrides of alkaline earth sub-nitrides were experimentally synthesized more than five decades ago.[39-41] Preliminary database screenings using Ca$_2$N as template yield 2 more nitrides and 8 carbides as layered electrides, nevertheless, all of them have the same Anti-CdCl$_2$ structure (R-3m).[42, 43]

While these endeavors by design or database screening yield some new layered electrides, the results are seriously narrowed down by the multiple presumptions applied. For example, all these works only focused on the binary A$_2$B and AB compounds. And the preliminary database screenings relied too much on the very limited knowledge of a single experimentally grown compound (Ca$_2$N). Hitherto, researches on experimentally grown layered electrides mainly focus on the nitrides and carbides with the same structure and similar interlayer 2D anionic electrons of Ca$_2$N.[44-46] One may ask if there exist other layered electride systems with completely different compositions, structures, and anionic electrons from Ca$_2$N. To answer this question, it is timely to carry out more exhaustive search and identify potentially useful electrides from experimentally grown compounds.

Here, to avoid missing potential candidates, we do not presume any specific structures, element groups, nor restrict compositions to any particular chemical formula. Instead, we apply only the basic descriptors directly derived from the definition of electrides, i.e., imbalanced total oxidation states and localized electrons at interstitial space. Structural stability is considered by limiting the candidates with total energies within a certain value above the energy Hull ($E_{Hull}$).[47] And we extend the screening pool from binary compounds used in previous researches to ternary. Removing these restrictions enable us to identify new unique layered electrides, such as the ones with one-dimensional (1D) anionic electrons trapped in connected tubes. Such complex structures (with tubes) are not possible for the simple A$_2$B or AB compounds with a small number of atoms per formula.

Our screening has successfully identified 12 new layered electrides, all of which are experimentally synthesizable and stable with small energies above the Hull. Completely new layered electride systems with 1D anionic electrons in the connected tubes and 0D magnetic anionic electrons at the center of metal-halogen hexagons are found. Remarkable coexistence of anionic electrons with different magnetic properties, localization or dimensions in one structure is first reported in this work. Our results have not only significantly increased the members of and provided new perspectives on experimentally grown layered inorganic electrides, but also display the importance of the careful choice of presumptions for exhaustive screenings.

METHODS

The procedure of high-throughput screening for layered electrides is shown in Figure 1. The screening starts from more than 67000 inorganic compounds in Materials Project database.[48] A geometry-based algorithm was used to find the possible layered materials in the database.[49] Then binary and ternary layered compounds with imbalanced total oxidization states are selected. Previous studies have shown that materials with $E_{hull}$ less than 50 meV/atom are reasonably synthesizable experimentally.[50, 51] Considering electrides can be metastable, we increase the threshold of $E_{hull}$ to 70 meV/atom. Actually, all of the new electrides screened here have lower $E_{hull}$ than 50 meV/atom, except Er$_6$I$_7$ with 62 meV/atom. Nevertheless, all of these materials have been experimentally grown before.[52]

The above steps lead to 498 candidates for further *ab initio* calculations. The standard high-throughput spin-polarized calculations with default input parameters developed by Materials Project were used with Vienna *ab initio* simulation package (VASP).[53-56] To better estimate the interlayer dispersion interactions in layered materials, the dispersion-corrected vdW-optB88 exchange-correlation functional was applied.[57-60] Electron localization functions (ELF) were calculated for all the materials. A simple algorithm on ELF for either spin-up or spin-down electrons was used to screen compounds with localized electrons (ELF threshold 0.75) in the interstitial area, and the ones with an significant ratio ω (>0.001) of volume between the interstitial localized electrons and the whole crystal cell were taken as possible electride candidates.[37] The distribution of ω all the 498 candidates can be found in the Supporting Information Figure S1. Most of them (>400) don't have localized electrons (ELF>0.75) at the interstitial space. A cut-off value of ω (0.1%), which is comparable with the known electride Ba$_2$N (0.15%),[42] is chosen to screen candidates for further consideration. We have carefully checked all these candidates with further examination of the ELF distribution, the size of the interstitial space, projected density of states (PDOS) contribution of pseudoatoms added at the sites of interstitial spaces,[61, 62] distribution of band decomposed charge dentistry, as well as literature review on the experimental results, and we exclude all the suspicious and hypothetical compounds. It is noted we define the Wigner-Seitz radii of interstitial sites by the subtraction of the distances be-tween interstitial sites and surrounding metal ions by the Wigner-Seitz radii of the metal ions used in VASP potential files to calculate PDOS. The details of these examinations can be found in the Supporting Information. To include the strong on site Coulomb interaction of 4f electrons, GGA+U was used for rare earth-elements.[63-65]

RESULTS AND DISCUSSION

The screening process in the Method section yields 24 most promising layered electrides and they can be easily classified into different groups based on their structure prototypes as shown in Table 1. Among these electrides, 12 of them with alkaline earth metal nitrides, rare earth metal carbides, rare earth metal halides, and early transition metal halides have been reported (the materials with light orange background in Table 1).[37, 42, 43] Successful screening of these structures demonstrate the reliability of the screening procedure applied in this work. The other 12 compounds are first identified as electrides here (the compounds in italic font with light blue background in Table 1), and have been experimentally grown before (see the ICSD id numbers[52] in the Supporting Information Table S1). The calculated lattice parameters agree well with experimental results. These comparisons, along with their calculated electronic properties and ELF distributions can be found in Supporting Information Table S1 and Fig. S3-S12.

These 12 compounds have large interstitial space, comparable or larger PDOS contribution from pseudoatoms at anionic electron sites to/than that from metal atoms, and band decomposed charge dentistry mainly distributing at interstitial sites (details can be found in Supporting Information). From the experimental side, although the direct measurements of the electride nature of these compounds are missing, hints from existing literature are there. For example, electrides can be taken as the stoichiometric deficient version of electron saturated counterpart materials, such as $Ca_2NX$ (X=Cl, Br) for $Ca_2N$,[66] and $Ca_{24}Al_{28}O_{66}$ for $Ca_{24}Al_{28}O_{64}$.[5] There are also sufficient examples for each group such as hydrides for ZrCl-like structures,[67] $LaBr_2H$,[68] $ThI_2O$[69] for $MoS_2$-like structures, $Y_2Cl_3N$,[70] $Sc_5Cl_8N$[71] for the quasi-2D structures and $Ba_2LiN$ could be taken as missing one LiN per formula from BaLiN. These results confirm that the localized electrons at interstitial site are more like anions rather than the hybridized orbitals from direct metal-metal bonding. The structural and electronic properties, as well as comparison with experimental results for each new electride can be found in the Supporting Information.

It is interesting to note that most of these 12 layered electrides have small exfoliation energies (around or less than 100 meV/atom) except for $Ba_2LiN$ (231 meV/atom) (see Supporting Information Table S1). However, the exfoliation energy of $Ba_2LiN$ is still comparable with $Ca_2N$ (261 meV/atom), the latter of which have already been experimentally exfoliated.[33] Therefore, all of these new layered electrides are promising to be exfoliated to advanced 2D materials.

In the following, we will mainly discuss three groups of the new electrides that are elusive in the previous designs/screenings and use one example for each group to discuss their properties: 1, quasi-1D structures requiring more complex composition and more atoms per formula than $A_2B$ and $AB$; 2, $MoS_2$-like structures with new $AB_2$ and $A_2B_5$ composition; 3, a ternary compound $Ba_2LiN$. At last, we will provide some empirical rules deduced from these new layered electrides, which may guide further studies on these fascinating materials.

As seen in Figure 2, the structures in quasi-1D group composed of layers with connected 1-D tubes. The infinite chains are composed of $A_4B_6.6e^-$ (A=Gd, Sc, Tb, Y, B=Cl, Br) tubes for single A octahedra with bonded B atoms or $A_6B_8.10e^-$ (A=Er, Sc, B=Cl, I) for double A octahedra with bonded B atoms. Interestingly, the even number of anionic electrons within these 1D tubes fully occupy bands to energy levels as deep as around 1.5 eV below Fermi level, leading to semiconducting behaviour. A similar case with even number of anionic electrons fully occupying energy bands with a band gap can be found for $La_8Sr_2(SiO_4)_6.4e^-$.[72] However, the overall electronic properties of these quasi-1D electrides depend on how they are connected. For the ones by sharing halogen atoms like $Y_2Cl_3$ and $Gd_2Br_3$, the semiconducting behaviour is preserved with 0.69 and 0.12 eV band gap respectively (see Figure 3, Supporting Information Table S1, ,S8 and related discussions). $Sc_5Cl_8$ and $Tb_5Br_8$ can be taken as $A_4B_6.AB_2$. While the $A_4B_6$ tubes do not contribute conductivity, the excess electrons from the $AB_2$ junction lead to metallic behaviour for $Sc_5Cl_8$ and $Tb_5Br_8$ (see Supporting Information Table S1, Figure S9, S12 and related discussions). A similar case can be found for $Sc_7Cl_{10}$, which can be taken as $Sc_6Cl_8.ScCl_2$ (It is noted that the 1D tubes in $Sc_7Cl_{10}$ are composed of double $Sc_6Cl_8$ octahedra). $Er_6I_7$ is conducting as the $Er_6I_8$ tubes are connected in a way that one Iodine atom per formula unit is missing, leading to another one excess electron.

To better understand these materials in quasi-1D group, we provide detailed structural and electronic properties of $Y_2Cl_3$ as an example (see Figure 3). The ELF within Y octahedron clearly shows the coexistence of two types of anionic electrons localized within the layer plane (see Figure 3b). The type 1 anionic elections are closer to the Y atoms while type 2 ones stay at the center of the tubes. The type 1 anionic electrons are more localized than type 2 ones as they survive in a higher ELF isovalue threshold. From the band structure and band decomposed charge density with different energy levels as shown in Figure 3c, d, and e, the type 1 anionic electrons are contributed by the bands around -0.5 eV while type 2 from the bands around -1.5 eV. This is consistent with PDOS shown in Figure 3a, which also indicates that the main contribution to these bands is from anionic electrons rather than the Y ions. The smaller bandwidth of type 1 electrons is another proof for their more localized nature than type 2 elec-trons. The high band dispersion along Γ-X and localization along Γ-Y direction are in line with the quasi-one-dimensional character of X2 anionic electrons as shown in Figure 3b and 3e. The compounds of this group have shown that the anionic electrons are not necessarily contributed by the bands around Fermi level, and they can have different localization even in the same structure.

MoS$_2$ is one of the most studied 2D materials, partially due to its rich physics with polymorphs.[73-76] The layered electrides in MoS$_2$-like structures show different phases of MoS$_2$ (see Figure 4). LaBr$_2$ have an H-phase structure, ThI$_2$ with a mixture of H and T-phase. La$_2$Br$_5$ has a variant H-phase with connected 1D honeycomb lattices. Interestingly, the anionic electrons can only be found in structures in H phase. For the H phase and its variant, the excess electrons in $[AB_2]^+(e^-)$ (A=La, B=Br) and $[A_2B_5]^+(e^-)$ (A=La, B=Br) are almost fully spin-polarized, contributing ~1 $\mu_B$ per formula. However, for the H-phase ThI$_2$, there are two excess electrons per formula and these systems are non-magnetic.

Figure 4d show the band structure with both nonspin-polarized and spin-polarized calculations, as well as PDOS of LaBr$_2$ as an example of the layered electrides in H and its variant phases. When a nonspin-polarized calculation is applied, the half-filled energy band of the one anionic electron per formula crosses the Fermi level, leading to a metallic property. Interestingly, a difference picture is shown when spin polarization is involved. The anionic electron band split into two bands with a spin-splitting gap of 0.3 eV, yielding localized magnetic anionic electrons. Such anionic electrons have been reported in previous studies, and was argued to be attributed to their half-filled and low-dimension nature, in line with the case of LaBr$_2$ (with half-filled and 0D anionic electron).[38,77]

The PDOS and the decomposed charge density (as shown in Figure 4c) of the valence band both indicate the electride nature of LaBr$_2$. The bandwidth for the entire spin-splitting bands is around 0.75 for LaBr$_2$. The small band gaps and large bandwidths make them good bipolar magnetic semiconductors.[78] By adjusting the Fermi level, half-metallicity with opposite spin polarization can be tuned for these materials. Thus, the magnetic layered electrides in this group are promising for spintronics applications. And the phase-dependent properties, which are possible to show phase transition under external stimuli, may be useful for applications like sensors.

Ba$_2$LiN has a completely new structure for electrides with infinite mutually perpendicular rows of edge-sharing Li-Ba$_5$N octahedra with symmetry P4$_2$/nmc. Different from the reported layered electrides,[36, 37, 42, 43] Ba$_2$LiN owns two different metal atoms, Ba and Li, both of which provide anionic electrons while in different distribution, yielding Ba$_2$LiN a unique electride with a coexistence of 1D and 0D anionic electrons. As shown in Figure 5a, the 0D anionic electrons (X1) is surrounded by 5 Ba atoms. And the Li ions are not uniformly distributed along the infinite rows, which leads to anionic electrons of difference size (X2 and X3). From the PDOS shown in Figure 5c, the occupied energy bands near Fermi level are mainly contributed by the anionic electrons. These bands of anionic electrons (in one spin state) are highlighted in bold green in Figure 5b. As shown in Figure 5 d-i, their decomposed charge densities are mainly localized on the interstitial sties, confirming the electride nature of Ba$_2$LiN.

The structure of binary ZrCl-like electrides is not new for electride[37], although their detailed studies are still missing. Our screening has discovered two new compounds TbBr and TbCl, and several related hydrides. Unfortunately, the H contents in these hydrides are not precisely determined, thus we do not consider them in this work. For the binary ZrCl-like compounds, a coexistence of magnetic and non-magnetic anionic electrons could be found (see Figure S3 and S4, and related discussions in Supporting Information). And their interesting relation with hydrides may be worthy of further studies.

At last, we perform statistical analysis on all the electrides screened in this work to provide some rules of thumb that could be applied for further researches. A statistics on the composed elements of these 24 layered electrides can be found in Figure 6a and 6b. Contradicted with the previous assumption that electride cations prefer s-block elements,[42] most (76%) layered electrides are composed of lanthanide or early transition metal as cations. There are only 16% of layered electrides with alkaline earth metals. And only one alkaline metal, that is, lithium is found in Ba$_2$LiN. On the anion side, halogens are highly preferred with a total composition of 62.50%. Carbides and nitrides contribute 20.83% and 16.67% of all the screened electrides here, respectively, most of which have been previously reported.[37, 42, 43] These results demonstrate that lanthanide and early transition metal halides are a unexplored area for layered electrides that worthy further research studies. It also reminds us that the previously reported electrides (nitrides and carbides) make up a small portion with some specific groups, and high-throughput screenings based on these known compounds might be biased.

Figure 6c shows the electronegativity of all the composed elements.[69] Cationic elements with low electronegativity may be more favourable to form electrides (with a highest value of 1.36 for Sc). For binary electrides, anionic elements with relatively high electronegativity seem to be preferable (with a lowest value of 2.55 for C). Generally speaking, electrides prefer compositions of relatively low-electronegativity cationic elements and high-electronegativity anionic elements.

In a previous work, a sufficiently large cation/anion size ratio was argued to be prerequisite for stable 2D electrides.[36] Our analysis shows supportive results for carbides and nitrides. However, an opposite trend could be found for layered halides with a maximum cation/anion size ratio of 0.53 (see Figure 6d). From these results, we could see: 1, halides are a new class of layered electrides with completely different properties with previous reported carbides and nitrides; 2, compared to the electronegativity, this empirical rule on cation/anion size depends on specific systems and no clear universal rule can be established for all the layered electrides.

On the structural side, this work has shown four possible distributions for anionic electrons: 1. in the 1D tubes (quasi-1D structures); 2. at the center of the in-plane metal triangles ($MoS_2$-like structures); 3. in the intralayer and between the cations layers (ZrCl-like structures and hydrides); 4. in the interlayer and between the cations layers (all the nitrides and carbides). Except the fourth case, the other three are less studied with the first two types of layered electrides first found in this work. It is noted this work aims to identify new prototypes for layered electrides, the individual materials may not be exhaustive due to comprehensiveness of databases and the criteria applied in the screening process.

CONCLUSIONS

In summary, we have performed an extensive screening to identify experimental synthesizable layered electrides. Removal the constraints applied in previous studies helps us find completed new layered electrides with unique structural, chemical, electronic and magnetic properties as well as different anionic electron distributions from known layered electrides. These new structures not only enrich the pool of experimental synthesizable layered electrides but also are promising to be exfoliated into advanced 2D materials.

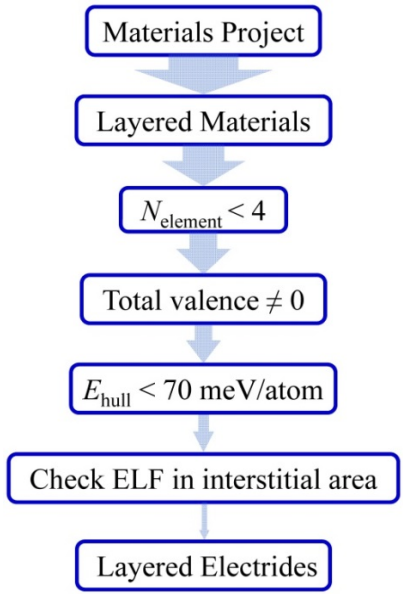

Figure 1. High-throughput screening procedure for layered electrides, starting from more than 67000 inorganic compounds in Materials Project database. The decreasing size of blue arrow indicates the number of candidates after each screening step.

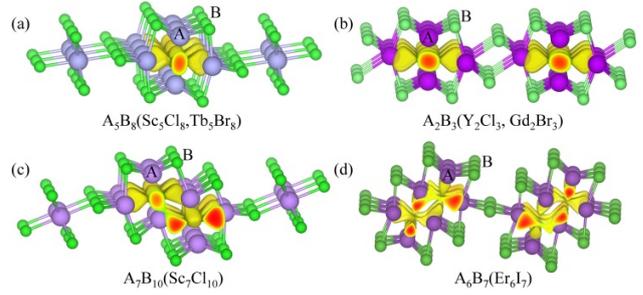

Figure 2. Structure prototypes in side view with ELF in yellow for layered electrides in quasi-1D structures (only one monolayer is shown). Formulas before brackets are compositions and the ones in brackets are respective materials within the same structure prototype. It is noted the specific structure parameters and ELF distribution for the materials in a same structure prototype can be slightly different. Their respective details can be found in Supporting Information.

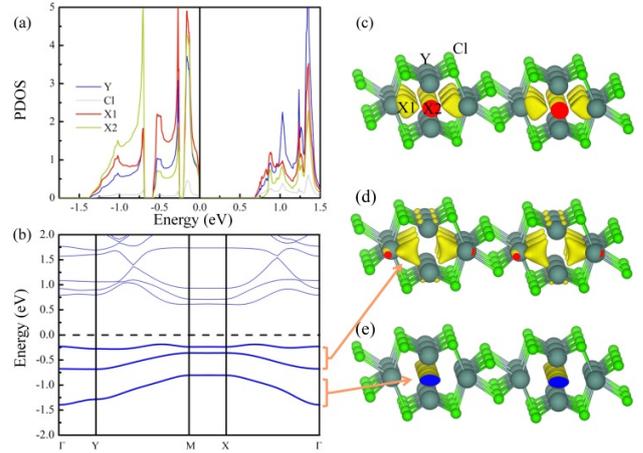

Figure 3. (a) The projected DOS with pesudoatoms at the sites of anionic electrons for $Y_2Cl_3$. (b) Band structure of $Y_2Cl_3$ monolayer. The anionic electron energy bands are highlighted by bold blue lines. The Fermi level is aligned to 0 eV. (c) Structural guidance for $Y_2Cl_3$ with ELF (isosurface value of 0.75) in yellow. X1, X2 denote the different anionic electrons. (d) and (e) are the band [green bold lines shown in (b)] decomposed charge densities in side view for $Y_2Cl_3$.

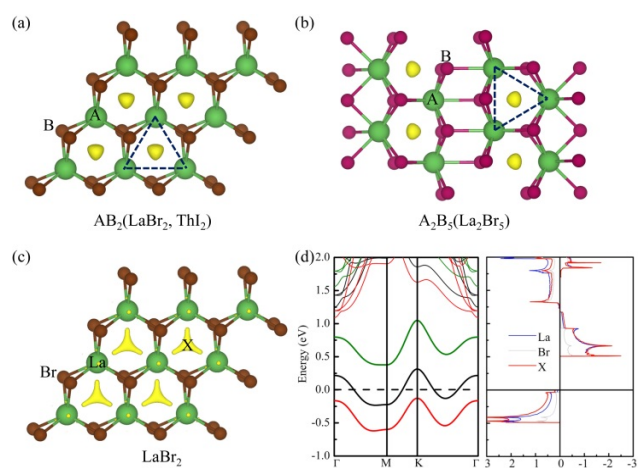

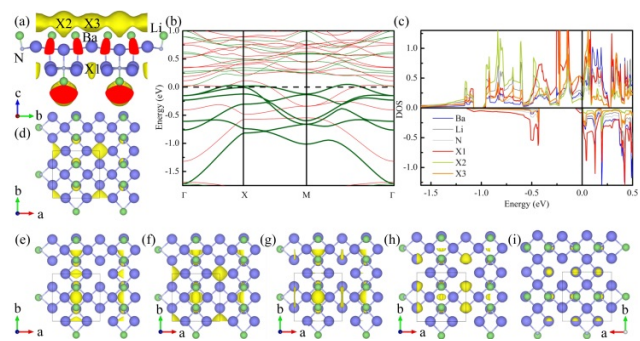

Figure 4. Electronic and structural properties (top view) of layered electrides in $MoS_2$-like structures (only one monolayer is shown). (a) $AB_2$ ($LaBr_2$, $ThI_2$) with anionic electrons localized at the centre of the metal-halogen hexagons. (b) $A_2B_5$ ($La_2Br_5$) with anionic electrons localized at the centre of the metal-halogen hexagons. (c) Decomposed charge density of the bold red band shown in (d) for $LaBr_2$ monolayer. (d) Band structure and PDOS with pesudoatoms at the site X shown in (c) for $LaBr_2$ monolayer. The nonspin-polarized bands are represented in black, while the red and green lines are the spin-up and spin-down bands respectively. The Fermi level is aligned to 0 eV.

Figure 5. Electronic and structural properties of $Ba_2LiN$. (a) Structural guidance for $Ba_2LiN$ in side view with ELF (isosurface value of 0.75) in yellow. X1, X2 and X3 denote the different anionic electrons. (b) Band structure of $Ba_2LiN$. (c) The projected DOS with pesudoatoms at the sites of anionic electrons for $Ba_2LiN$. (d), (e), (f), (g), (h) and (i) are the band [green bold lines shown in (b)] decomposed charge densities in top view for $Ba_2LiN$ in the order of lower to higher energies.

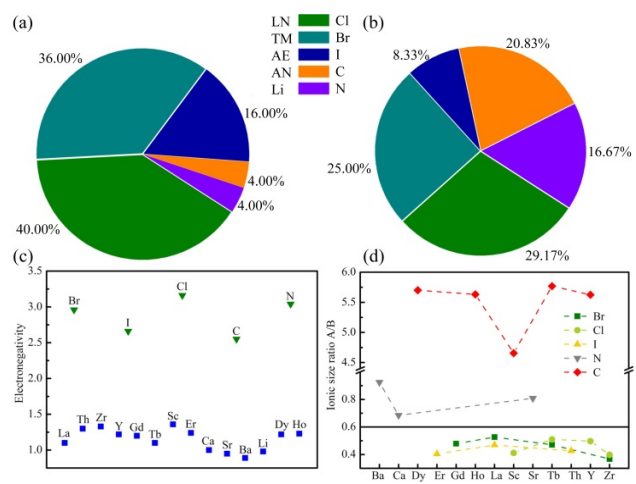

Figure 6. Statistic information for the 24 layered electrides screened in this work. (a) and (b) are the cation (left panel) and anion (right panel) distribution in percentage. LN, TM, AE, and AN stand for lanthanide, early transition metal, alkaline earth metal and actinide respectively, while others are elements in the chemical symbol. Noted that Li only exists in a ternary compound of $Ba_2LiN$. c) Electronegativity[69] of the compositing elements of the 24 layered electride. d) Shannon ionic size ratio of A (cations) and B (anions) for the binary electrides screened in this work.

Table 1. The 24 layered electrides screened in this work, which can be classified based on their structure types into prototype structures of ZrCl, $MoS_2$, pseudo-1D, anti-$CdCl_2$. [a]

| ZrCl | *TbCl* | *TbBr* | ScCl | YCl | ZrCl | ZrBr |   |
|---|---|---|---|---|---|---|---|
| $MoS_2$ | *$LaBr_2$* | *$ThI_2$* | *$La_2Br_5$* |   |   |   |   |
| Quasi-1D | *$Y_2Cl_3$* | *$Gd_2Br_3$* | *$Sc_5Cl_8$* | *$Tb_5Br_8$* | *$Er_6I_7$* | *$Sc_7Cl_{10}$* |   |
| Anti-$CdCl_2$ | $Ca_2N$ | $Sr_2N$ | *$Ba_2N$* | $Y_2C$ | $Sc_2C$ | $Tb_2C$ | $Dy_2C$ $Ho_2C$ |
| others | *$Ba_2LiN$* |   |   |   |   |   |   |

[a]The 12 materials listed in italic fond with light blue background are new electrides identified in this work while the other 12 with light orange background were reported previously.

## ASSOCIATED CONTENT

**Supporting Information**.
This material is available free of charge via the Internet at http://pubs.acs.org."
Basic structural, electronic and magnetic properties, as well as the distribution of anionic electrons of the 12 layered electrides

## AUTHOR INFORMATION

**Corresponding Author**
* E-mail: phyfyp@nus.edu.sg

**Author Contributions**

J.Z. performed the calculations with assistance from L.S., M.Y. and Y.P.F. All authors contribute to the discussion and the writing of the manuscript. Y.P.F supervised the project.

**Funding Sources**

This work was supported by AcRF Tier 1 Research Project (R-144-000-361-112).

**Notes**

The authors declare no competing financial interest.

## ACKNOWLEDGMENT

We acknowledge Centre for Advanced 2D Materials and Graphene Research Center at National University of Singapore, National Supercomputing Centre Singapore for providing computing resource.